\documentstyle[11pt,newpasp,twoside]{article}

\begin{document}

\title{Modelling galaxy clustering at high redshift}
\author{Eelco van Kampen}
\affil{Institute for Astronomy, University of Edinburgh, Royal Observatory,
Blackford Hill, Edinburgh EH9 3HJ, United Kingdom}

\begin{abstract}
Most phenomenological galaxy formation models show a discrepancy
between the predicted Tully-Fisher relation and the luminosity
function. We show that this is mainly due to overmerging of galaxy haloes,
which is inherent in both the Press-Schechter formalism and dissipationless
N-body simulations. This overmerging problem be circumvented
by including a specific galaxy halo formation recipe into an
otherwise standard N-body code. Resolving the overmerging
also allows us to include models for chemical evolution and
starbursts, which improves the match to observational data {\it and}
renders the modelling more realistic.
We use high-redshift clustering data to try and distinguish models
which predict similar results at low redshifts for different sets
of parameters. 
\end{abstract}

\keywords{cosmology: theory - dark matter - large-scale structure of Universe
- galaxy formation}

\section{Introduction}

There has been significant recent progress in the study of galaxy
formation within a cosmological context, mainly due to a phenomenological
approach to this problem. The idea is to start with
a structure formation model that describes where and when galactic
dark haloes form. A simple description of gas dynamics and star formation
provides a means to calculate the amount of stars forming in these haloes.
Stellar population synthesis models then provide the spectral evolution,
i.e.\ luminosities and colours, of these galaxies.

Many physical processes are modelled as simple functions of the circular
velocity of the galaxy halo. Therefore, the Tully-Fisher relation is
the most obvious observational relation to try and predict, as it
relates the total luminosity of a galaxy to its halo circular
velocity. However, most phenomenological galaxy formation models do
not simultaneously fit the I-band Tully-Fisher relation and the
B or K band luminosity function. When one sets the model parameters
such that the Tully-Fisher relation has the right normalization, the
luminosity functions generally overshoot (e.g.\ Kauffmann, White \&
Guiderdoni 1993; Kauffmann, Colberg, Diaferio \& White 1999), certainly
for the $\Omega=1$, $H_0=50$ km s$^{-1}$ Mpc$^{-1}$
standard CDM cosmology (in the form given by Davis et al.\ 1985) that
we consider in this paper. Alternatively, when making sure that the
luminosity functions matches by changing some of the model parameters,
the Tully-Fisher relation ends up significantly shifted with respect to
the observed relation (e.g.\ Cole et al.\ 1994; Heyl et al.\ 1995).

In order to keep the modelling as analytical as possible, an extension
to the Press \& Schechter (1974) prescription for the evolution of galaxy
haloes (e.g.\ Bond et al. 1991; Bower 1991; Lacey \& Cole 1993; Kauffmann
\& White 1993) has been a popular ingredient for implementations of a
phenomenological theory of galaxy formation. 
However, the EPS formalism is designed to identify collapsed
systems, irrespective of whether these contain surviving subsystems.
This `overmerging' of subhaloes into larger embedding haloes is
relevant to the problem of matching both the galaxy luminosity
function and the Tully-Fisher relation, as the central galaxy in an
overmerged halo is the focus of a much larger cooling gas reservoir
than the reservoir that galaxy is to focus of in case its parent
subhalo survives.
Traditional N-body simulations suffer from a similar overmerging
problem (e.g.\ White 1976), which is of a purely numerical nature,
caused by two-body heating in dense environments when the mass
resolution is too low (Carlberg 1994; van Kampen 1995).

In order to circumvent these problems,
we use an N-body simulation technique that includes a built-in
recipe for galaxy halo formation, designed to prevent overmerging
(van Kampen 1995, 1997), to generate the halo population and its
formation and merger history. This resolves most of the discrepancy
sketched above, {\it and}\ allows us to make the modelling more
realistic by adding chemical evolution and a merger-driven bursting
mode of star formation to the modelling.
Once stars are formed, we apply the stellar population synthesis models
of Jimenez et al.\ (1998) to follow their evolution. We have enhanced
these models with a model for the evolution of the average metallicity
of the population, which depends on the starting metallicity.
Feedback to the surrounding material means that cooling properties of
that material will change with time, affecting the star formation rate,
and thus various other properties of the parent galaxy.

\section{Overview of the phenomenological model}

The key ingredients of the model are described below. We refer
to van Kampen et al.\ (1999) for a much more detailed description
and discussion of the model, and a list of the choices for the various
parameters involved.

\subsubsection
{\it The merging history of dark-matter haloes.}
This is often treated by Monte-Carlo realizations of the
analytic `extended Press-Schechter' formalism, which ignores
substructure. We use a special N-body technique 
to prevent galaxy-scale haloes undergoing `overmerging'
owing to inadequate numerical resolution.

\subsubsection
{\it The merging of galaxies within dark-matter haloes.}
Each halo contains a single galaxy at formation. When haloes 
merge, a criterion based on dynamical friction is used to decide how
many galaxies exist in the newly merged halo. The most massive
of those galaxies becomes the single central galaxy to which
gas can cool, while the others become its satellites. 

\subsubsection
{\it The history of gas within dark-matter haloes.}
When a halo first forms, it is assumed to have
an isothermal-sphere density profile. A fraction
$\Omega_b/\Omega$ of this is in the form of gas
at the virial temperature, which can cool to form
stars within a single galaxy at the centre of the halo.
Application of the standard radiative cooling
curve shows the rate at which this hot gas cools
below $10^4$~K, and is able to form stars.
Energy output from supernovae reheats some of the
cooled gas back to the hot phase. When haloes
merge, all hot gas is stripped and ends up in the new halo.

\subsubsection
{\it Quiescent star formation.}
The star formation rate is equal the ratio of the amount of
cold gas available and the star-formation timescale.
The amount of cold gas available depends on the merger
history of the halo, the star formation history, and
the how much cold gas has been reheated by feedback
processes.

\subsubsection
{\it Starbursts.}
We also model star bursts, i.e.\ the star-formation rate may
suffer a sharp spike following a major merger event. 

\subsubsection
{\it Feedback from star-formation.}
The energy released from young stars heats cold
gas in proportion to the amount of star-formation,
returning it to the reservoir of hot gas.

\subsubsection
{\it Stellar evolution and populations.}
Our work assumes the spectral models of Jimenez et al.\ (1998);
for solar metallicity, the results are not greatly
different from those of other workers.
The IMF is generally taken to be Salpeter, but any choice is possible.
Unlike other workers, we take it as established that the
population of brown dwarfs makes a negligible contribution
to the total stellar mass density, and we do not
allow an adjustable $M/L$ ratio, $\Upsilon$, for the stellar
population.

\subsubsection
{\it Chemical evolution.}
The evolution of the metals must be followed, for two reasons:
(i) the cooling of the hot gas depends on metal content;
(ii) for a given age, a stellar population of high metallicity will
be much redder. The models of Jimenez et al.\ (1998) allow
synthetic stellar populations of any metallicity to be
constructed. 

\section{Low-redshift results}

With the set-up described above we match both the B and K band
luminosity function and the I-band Tully-Fisher relation, for an
$\Omega=1$ standard CDM structure formation scenario. Resolving the
overmerging problem is the major contributor to this result,
but the inclusion of chemical evolution and starbursts are also
important ingredients.

The new ingredients we have added to the modelling of galaxy formation
are needed in order to make the models more realistic, and are not
introduced simply in order to give yet more free parameters. Nevertheless,
our resolution to the Tully-Fisher / luminosity function discrepancy
may well not be unique, and various other changes to the ingredients of the
phenomenological galaxy formation recipe might produce similar
results. For example, we have not studied the influence cosmological
parameters have on the model galaxy populations, where $\Omega$,
$\Lambda$, and $\sigma_8$ are likely to be the important parameters.
Other types of ingredients are possible as well:
Somerville \& Primack (1998) resolve some of the discrepancy using
a dust extinction model plus a halo-disk approach to feedback.

\section{High-redshift clustering}

One way of resolving the worries about degeneracies in the
cosmological/physical parameter space will be to include data at
intermediate and high redshifts, which is being gathered with
increasing speed and ease, and at increasingly higher redshifts.
In this contribution we show a preliminary comparison of the
correlation properties of galaxies at redshift $z=3$. Recently,
Giavalisco et al.\ (1998) gave an estimate for the galaxy-galaxy
correlation function $\xi(r)=(r_0/r)^\gamma$ for a sample of
Lyman-break galaxies at this redshift. They found $r_0=2.1 h^{-1}$Mpc
and $\gamma=2.0$. 

We selected our model galaxies in exactly the same way as Giavalisco
et al.\ (1998) did, and compared two of the models produced by van Kampen
et al.\ (1999), models $n$ and $b$, to the observational data.
The first model ($n$), which is as close as possible to the model
by Cole et al.\ (1994), but with the mass-to-light parameter $\Upsilon=1$,
gives $r_0=3.5 h^{-1}$Mpc and $\gamma=1.72$. 
The second model ($b$), which includes starbursts and chemical evolution,
gives $r_0=4.4 h^{-1}$Mpc and $\gamma=2.1$. Both models fit the
correlation function at $z=0$ very well, and cannot be distinguished
from each other.

As the observed correlation data are still relatively uncertain at this
moment in time, it is premature to rule out models on the basis of this
data. The two models discussed above have similar predictions
for low redshifts, but predict different clustering properties at
high-redshift. However, the differences are not large, so one needs
either really good data, or a much larger variety of observational
characteristics of the high-redshift galaxy population.

\acknowledgments
Many thanks for the loud vocal support from outside the conference
building during my presentation. I like to think that the people of Marseille
just wanted to show how much they supported everything I said ...

\end{document}